\documentclass[reqno,11pt,a4paper]{amsart}
\pdfoutput=1
\usepackage[margin=2.3cm,footskip=30pt]{geometry}
\usepackage[ascii]{inputenc}
\usepackage{etoolbox}
\usepackage{silence}
\usepackage[dvipsnames]{xcolor}
\usepackage[bookmarksnumbered=true,linktocpage=true,hypertexnames=false,colorlinks=true,linkcolor=blue,urlcolor=BrickRed,citecolor=ForestGreen,anchorcolor=green,breaklinks=true,pdfusetitle]{hyperref}
\usepackage{bbm,braket,mathrsfs,amsmath,amssymb,amsthm,amsfonts,latexsym,mathtools,graphicx,enumitem,caption,booktabs,bm,mathdots,ellipsis,mleftright}
\usepackage{microtype}
\usepackage[T1]{fontenc}
\usepackage{textcomp}
\usepackage[english]{babel}
\usepackage[capitalize,nameinlink]{cleveref}
\numberwithin{equation}{section}
\allowdisplaybreaks[4]
\setenumerate{label=(\roman*)}
\newtheorem{thm}{Theorem}\crefname{thm}{Theorem}{Theorems}
\crefname{lem}{Lemma}{Lemmas}
\newtheorem{cor}[thm]{Corollary}\crefname{cor}{Corollary}{Corollaries}
\crefname{rem}{Remark}{Remarks}
\crefname{prop}{Proposition}{Propositions}
\crefname{conj}{Conjecture}{Conjectures}
\crefname{defn}{Definition}{Definitions}
\DeclareMathOperator{\Hom}{Hom}
\DeclareMathOperator{\GL}{GL}
\DeclareMathOperator{\Sym}{Sym}
\DeclareMathOperator{\KT}{KT}
\DeclareMathOperator{\outarrows}{out}
\DeclareMathOperator{\inarrows}{in}
\renewcommand{\vec}{\bm}
\newcommand{\C}{\mathbbm C}
\newcommand{\R}{\mathbbm R}
\newcommand{\Z}{\mathbbm Z}
\newcommand{\N}{\mathbbm N}
\newcommand{\Q}{\mathbbm Q}
\newcommand{\CV}{\vec V}
\newcommand{\CH}{\mathcal H}
\newcommand{\op}{\oplus}
\newcommand{\ot}{\otimes}

\DeclarePairedDelimiter\parens{\lparen}{\rparen}

\begin{document}
\title{Moment cone membership for quivers in strongly~polynomial~time}
\author{Mich\`ele Vergne}
\address{Mich\`ele Vergne: Universit\'e Paris 7 Diderot, Institut Math\'ematique de Jussieu, Sophie Germain, case 75205, Paris Cedex 13, France}
\email{michele.vergne@imj-prg.fr}
\author{Michael Walter}
\address{Michael Walter, Faculty of Computer Science, Ruhr University Bochum, Universit\"atsstr.~150, 44801 Bochum, Germany}
\email{michael.walter@rub.de}

\begin{abstract}
In this note we observe that membership in moment cones of spaces of quiver representations can be decided in strongly polynomial time, for any acyclic quiver.
This generalizes a recent result by Chindris-Collins-Kline for bipartite quivers.
Their approach was to construct ``multiplicity polytopes"  with a geometric realization similar to the Knutson-Tao polytopes for tensor product multiplicities.
Here we show that a less geometric but straightforward variant of their construction leads to such a multiplicity polytope for any acyclic quiver.
Tardos' strongly polynomial time algorithm for combinatorial linear programming along with the saturation property then implies that moment cone membership can be decided in strongly polynomial time.
The analogous question for semi-invariants remains open.
\end{abstract}

\maketitle

\section{Introduction}\label{sec:intro}
Let $Q=(Q_0,Q_1)$ be a quiver, where~$Q_0$ is the finite set of vertices and~$Q_1$ the finite set of arrows.
We use the notation~$a:x\to y$ for an arrow~$a\in Q_1$ from~$x\in Q_0$ to~$y\in Q_0$.
We allow~$Q$ to have  multiple arrows between two vertices, but no cycles.
For simplicity we assume that~$Q$ is connected. 
A \emph{dimension vector} for~$Q$ is a family~${\bf {n}} = (n_x)_{x\in Q_0}$ of non-negative integers.
Given such a dimension vector, we consider the family of complex vector spaces~$\CV=(V_x)_{x\in Q_0}$ with~$V_x=\C^{n_x}.$
Then the complex vector space of \emph{representations} of the quiver~$Q$ on~$\CV$ is defined by
\[ \CH_Q(\CV) \coloneqq \bigoplus_{a:x\to y\in Q_1} \Hom(V_x, V_y). \]
The Lie group~$\GL_Q({\bf{n}}) = \prod_{x\in Q_0} \GL(n_x)$  acts naturally on $\CH_Q(\CV)$, and so acts naturally on the space $\Sym^*(\CH_Q(\CV))$ of polynomial functions on~$\CH_Q(\CV)$.
Decompose
\[ \Sym^*(\CH_Q(\CV))=\bigoplus_{\bm{\lambda}} m_Q(\bm{\lambda}) V^{\bm{\lambda}}, \]
where $V^{\vec\lambda}$ denotes the irreducible representation of $\GL_Q({\bf n})$ with highest weight $\bm\lambda=(\lambda_x)_{x\in Q_0}$.
The assumption that $Q$ has no cycles implies that the multiplicities $m_Q(\bm{\lambda})$ are finite.
The associated \emph{moment cone} is defined as the polyhedral convex cone generated by the~$\vec\lambda$ such that~$m_Q(\vec\lambda)>0$.

When~$Q$ is the quiver
\begin{equation}\label{eq:quiver}
  1 \rightarrow 3 \leftarrow 2
\end{equation}
and the dimension vector is $\bm n=[n,n,n]$, then $m_Q({\lambda_1,\lambda_2,\lambda_3})$ is the multiplicity of the trivial representation of $\GL(n)$ in $V^{\lambda_1}\otimes V^{\lambda_2}\otimes V^{\lambda_3}$, that is, the Clebsch-Gordan or \emph{Littlewood-Richardson coefficient}~$C(\lambda_1, \lambda_2,\lambda_3)$.%
\footnote{In the literature these are often defined in an asymmetrical way as the multiplicity of $ V^{\lambda_3}$ in $V^{\lambda_1}\otimes V^{\lambda_2}$, i.e., as $c_{\lambda_1,\lambda_2}^{\lambda_3} := C(\lambda_1, \lambda_2,\lambda_3^*)$, with $\lambda_3^*$ the highest weight of the dual representation.}
Knutson-Tao~\cite{knutson1999honeycomb} constructed a family of polytopes~$\KT(\lambda_1,\lambda_2,\lambda_3)$, parameterized by $(\lambda_1,\lambda_2,\lambda_3) \in (\R^n)^3$, in an~$O(n^2)$-dimensional space, such that, when $(\lambda_1,\lambda_2,\lambda_3)$ is a dominant weight, the number of integral points in the polytope equals the Littlewood-Richardon coefficient~$C(\lambda_1,\lambda_2,\lambda_3)$.
Moreover, the polytope is given by a combinatorial linear program%
\footnote{Recall that a linear program~$Ax \leq b$ is called \emph{combinatorial} if the coefficients of~$A$ are polynomially bounded in the dimension of the problem~\cite{tardos1986strongly}.}
of the form~$A x \leq b(\lambda_1,\lambda_2,\lambda_3)$, where~$A$ depends only on the dimension~$n$ and~$b$ is a linear function of~$\lambda_1,\lambda_2,\lambda_3$ (in fact, the coefficients of~$A$ and~$b$ are in $1,0,-1$), see \cref{sec:preliminaries}.
Furthermore, they prove the \emph{saturation conjecture}: one has $C(\lambda_1,\lambda_2,\lambda_3)>0$ if and only if there exists a positive integer~$N$ such that~$C(N\lambda_1,N\lambda_2,N\lambda_3)>0$.
In particular, $C(\lambda_1,\lambda_2,\lambda_3)>0$ if and only if $\KT(\lambda_1,\lambda_2,\lambda_3)$ is nonempty, which then is also equivalent to membership in the associated \emph{moment cone}, which in the present situation means that there exist Hermitian matrices~$A_1,A_2,A_3$ with eigenvalues~$\lambda_1,\lambda_2,\lambda_3$, respectively, such that $A_1+A_2+A_3=0$ (Horn's problem).
Since $\KT(\lambda_1,\lambda_2,\lambda_3)$ is given a by a combinatorial linear program, it follows readily that all these problems can be decided by an algorithm that runs in strongly polynomial time%
\footnote{Recall that a \emph{strongly polynomial time} algorithm is one that performs a number of elementary arithmetic operations that is bounded by a polynomial in \emph{the number of} input numbers and which, considered in the Boolean or Turing model, is a polynomial space algorithm~\cite{grotschel2012geometric}.}
by using Tardos' algorithm~\cite{tardos1986strongly}, as was pointed out by Mulmuley-Narayanan-Sohoni~\cite{mulmuley2012geometric}.%
\footnote{We remark that there is a long line of works on computing moment polytopes, from a structural or algorithmic point of view. See, e.g., \cite{berenstein2000coadjoint,ressayre2010geometric,vergne2017inequalities,burgisser2017membership,burgisser2019towards} and references therein. While the membership problem in a natural and general setting is known to be contained in the complexity classes NP and coNP~\cite{burgisser2017membership}, no general algorithms are known that run in polynomial time in all parameters.}

This raises the natural question whether one can construct for \emph{any} quiver~$Q$ and dimension vector~$\vec n$ a polytope~$P_Q(\bm\lambda)$, given by a combinatorial linear program as above, such that the number of integral points in~$P_Q(\bm\lambda)$ is equal to the multiplicity~$m_Q(\vec\lambda)$.
Since the multiplicities are also saturated by results of Derksen-Weyman~\cite{derksen2000semi}, namely that~$m_Q(\vec{\lambda})>0 $ if and only if there exists an integer~$N>0$ with~$m_Q(N\vec{\lambda})>0$,%
\footnote{This was pointed out to us by Nicolas~Ressayre. See \cite[Section~8]{baldoni2019horn} for details.}
this would also imply obtain a strongly polynomial time algorithm for moment cone membership.
In a recent article, Chindris-Collins-Kline~\cite{chindris2022membership} achieved this in the case when~$Q$ is a \emph{bipartite} quiver with the same number of edges between any pair of source and sink vertices (Eq.~\eqref{eq:quiver} is such a quiver).
Their polytope has a concrete geometric description and is very similar to a Knutson-Tao polytope.
In this note we give a construction that is less geometric but applies to any quiver without cycles.

\begin{thm}\label{thm}
For any quiver~$Q$ and dimension vector~$\vec n$, there exists a family of polytopes~$P_Q(\vec\lambda)$ such that, when~$\vec\lambda$ is a dominant weight, the number of integral points in~$P_Q(\vec\lambda)$ equals $m_Q(\vec\lambda)$.
Moreover, $P_Q(\vec\lambda)$ can be described by a combinatorial linear program that can be generated in strongly polynomial time given~$Q$ (given by the number of vertices and the list of arrows, encoded by pairs of integers) and~$\vec\lambda$ (given by a list of integer vectors~$\lambda_x$ of size~$n_x$); the right-hand side of the inequalities depend linearly on~$\vec\lambda$ and all coefficients are in $\{0,1,-1\}$.
\end{thm}

We prove \cref{thm} in \cref{sec:construction}.
Since $P_Q(\vec\lambda)$ is a rational polytope, it is nonempty if and only if a $N P_Q(\vec\lambda)$ contains an integer point for some integer~$N>0$.
By construction, $P_Q(N \vec\lambda) = N P_Q(\vec\lambda)$ for all dominant~$\vec\lambda$ and~$N>0$.
Thus the saturation property implies that $m_Q(\vec\lambda)>0$ if and only if the polytope~$P_Q(\lambda)$ is nonempty.
Thus, as in~\cite{mulmuley2012geometric,chindris2022membership} one can use Tardos' algorithm~\cite{tardos1986strongly} to decide in strongly polynomial time whether~$m_Q(\vec\lambda)>0$ or, equivalently, whether~$\vec\lambda$ is in the corresponding moment cone.
Thus, \cref{thm} implies the following:

\begin{cor}\label{cor}
There exists a strongly polynomial time algorithm that decides membership in the moment cone when given as input a quiver~$Q$ and a dominant weight~$\vec\lambda$.
\end{cor}

Given a quiver~$Q$ and a dimension vector~$\vec n$, one can also consider the cone~$\Sigma_Q(\vec n)$ of weights of semi-invariant polynomials for this data.
That is, $\vec\sigma \in \Sigma_Q(\vec n)$ if and only if there exists a non-zero semi-invariant polynomial on the space~$\CH_Q(\CV)$ of weight~$\vec\sigma \in \Z^{Q_0}$.
Chindris-Collins-Kline asked if one can give a strongly polynomial time algorithm for the problem of deciding membership in~$\Sigma_Q(\vec n)$, which they called the \emph{generic semi-stability problem}~\cite[Problem~2]{chindris2022membership}.
Interestingly, \cref{cor} does \emph{not} imply a (strongly or not) polynomial time algorithm for this problem in the natural input model where~$\vec n$ and~$\vec \sigma$ are specified by lists of integer vectors of length~$\lvert Q_0\rvert$ each.
We discuss this in \cref{sec:semi} and also give some positive examples.

\section{Preliminaries}\label{sec:preliminaries}
Before giving proofs of the above results we first recall some notation.
For any integer~$n$, we label the dominant weights for~$\GL(n)$ by non-increasing sequences of integers~$\alpha\in\Z^n$.
We denote the corresponding irreducible representation of~$\GL(n)$ by~$V^\alpha$; it is polynomial if $\alpha_n\geq0$.
We abbreviate~$D(n) = \{ x \in \R^n : x_1 \geq \dots \geq x_n \geq 0 \}$, so that $D(n) \cap \Z^n$ are precisely the dominant weights corresponding to polynomial irreducible representations.

Next, we recall a description of the \emph{Knutson-Tao polytopes}:
For~$\lambda_1,\lambda_2,\lambda_3\in\R^n$,
\begin{equation}\label{eq:kt}
\begin{aligned}
  \KT(\lambda_1,\lambda_2,\lambda_3) = \Bigl\{ p\in W_n \;:\; &E_r(p) \geq 0 \quad \forall r=1,\dots,\tfrac{3n(n-1)}2, \\
  &B_k(p,\lambda_1,\lambda_2,\lambda_3)=0 \quad \forall k=1,\dots,3n, \\
  &S(\lambda_1,\lambda_2,\lambda_3)=0 \Bigr\},
\end{aligned}
\end{equation}
where $W_n = \R^{(n+1)(n+2)/2}$, the~$E_r$ are linear forms on~$W_n$ defining the so-called ``rhombus inequalities'', where the~$B_k$ are linear forms on $W_n \op (\R^n)^{\op3}$ defining the ``boundary equations'', and $S(\lambda_1,\lambda_2,\lambda_3)=\sum_{i=1}^n\lambda_{1,i} +\sum_{i=1}^n\lambda_{2,i} +\sum_{i=1}^n\lambda_{3,i}$ defines the so-called ``compatibility equation''.
The linear forms $E_r$, $B_k$, and $S$ all have coefficients in $\{0,1,-1\}$.
See, e.g., \cite{buch2000saturation}.
The list of inequalities can be generated in strongly polynomial time given~$\lambda_1,\lambda_2,\lambda_3$ as lists of integers.
As discussed in \cref{sec:intro}, if~$\lambda_1,\lambda_2,\lambda_3$ are dominant weights then the number of integral points in~$\KT(\lambda_1,\lambda_2,\lambda_3)$ equals the Littlewood-Richardson coefficient~$C(\lambda_1,\lambda_2,\lambda_3)$, defined as the dimension of the~$\GL(s)$-invariant subspace of~$V^{\lambda_1}\otimes V^{\lambda_2}\otimes V^{\lambda_3}$:
\begin{align}\label{eq:C vs KT}
  C(\lambda_1,\lambda_2,\lambda_3) = \# \parens*{ \KT(\lambda_1, \lambda_2, \lambda_3) \cap \Z^{(n+1)(n+2)/2} }
\end{align}
Note that~$C(\lambda_1^*,\lambda_2,\lambda_3)$ can also be interpreted as the multiplicity of~$V^{\lambda_1}$ in the tensor product~$V^{\lambda_2}\otimes V^{\lambda_3}$.

Finally, we recall the \emph{Cauchy formula}, which states that with respect to $\GL(n_x) \times \GL(n_y)$ we have the decomposition
\begin{align}\label{eq:cauchy general}
  \Sym^*(\Hom(\C^{n_x}, \C^{n_y})) \cong \!\!\!\!\!\!\bigoplus_{\mu \in D(n_x,n_y) \cap \Z^{n_x}}\!\!\!\!\!\! V^{\mu} \ot V^{d(\mu)^*},
\end{align}
where $D(n_x,n_y)$ denotes the set of vectors~$x \in D(n_x)$ such that $x_j = 0$ for all $n_y < j \leq n_x$ (this condition is vacuous when $n_x \leq n_y$),
and where~$d(\mu) \in \R^{n_y}$ denotes the vector obtained as follows: if $n_y \leq n_x$ then take the first~$n_y$ coordinates of~$\mu$, while if~$n_y > n_x$~extend~$\mu$ by zeros.
See, e.g., \cite[Eq.~(18) in \S{}8.3]{fulton1997young}.

\section{The construction}\label{sec:construction}
We first describe the construction for a simple example and then in the general situation.

\subsection{An example}
Consider the following quiver~$Q$,
\begin{equation}\label{eq:quiver4}
\begin{aligned}
  &\qquad ~2 \qquad \\
  &\nearrow ~\qquad \searrow \\
  1 & \qquad\qquad\quad 4 \; , \\
  &\searrow ~\qquad \nearrow \\
  &\qquad ~3 \qquad
\end{aligned}
\end{equation}
which is bipartite, but \emph{not} of the form considered in \cite{chindris2022membership}.
For simplicity, consider the dimension vector~$\vec n = [n,n,n,n]$.
For $V_x = V_y = \C^n$, the Cauchy formula in \cref{eq:cauchy general} states that
\[
  \Sym^*(\Hom(V_x, V_y)) \cong \bigoplus_{\mu \in D(n) \cap \Z^n} V^\mu \ot V^{\mu^*}
\]
with respect to $\GL(n) \times \GL(n)$, since in this case~$D(n,n) = D(n)$ and~$d(\mu)=\mu$.
Since it holds that~$\Sym(A \op B) = \Sym(A) \ot \Sym(B)$, we obtain the following decomposition with respect to the action of~$\GL_Q(\CV)$:
\[
  \Sym^*(\mathcal H_q(\CV))
\cong \bigoplus_{\alpha,\beta,\gamma,\delta \in D(n) \cap \Z^n} V_{1}^\alpha \ot V_{2}^{\alpha^*} \ot V_{1}^\beta \ot V_{3}^{\beta^*} \ot V_{2}^\gamma \ot V_{4}^{\gamma^*} \ot V_{3}^\delta \ot V_{4}^{\delta^*}
\]
where the factor $\GL(n_x)$ of $\GL_Q(\CV)$ associated with vertex~$x$ acts diagonally on all tensor factors with subscript~$x$.
By definition of the Littlewood-Richardson coefficients, it follows that
\begin{equation}\label{eq:m1234}
  m_Q(\lambda_1,\lambda_2,\lambda_3,\lambda_4)
= \sum_{\alpha,\beta,\gamma,\delta \in D(n) \cap \Z^n}
C(\lambda_1^*, \alpha, \beta)
C(\lambda_2^*, \alpha^*, \gamma)
C(\lambda_3^*, \beta^*, \delta)
C(\lambda_4^*, \gamma^*, \delta^*).
\end{equation}
We claim can write this multiplicity as the number of integral points in the following polytope:
\begin{align*}
  & P_Q(\lambda_1,\lambda_2,\lambda_3,\lambda_4)
= \Bigl\{
  (\alpha,\beta,\gamma,\delta,p_1,p_2,p_3,p_4) \in (\R^n)^{\op 4} \op W_n^{\op 4} \;:\; \alpha, \beta, \gamma, \delta \in D(n), \\
 &\quad p_1 \in \KT(\lambda_1^*, \alpha, \beta),
 p_2 \in \KT(\lambda_2^*, \alpha^*, \gamma),
 p_3 \in \KT(\lambda_3^*, \beta^*, \delta),
 p_4 \in \KT(\lambda_4^*, \gamma^*, \delta^*)
\Bigr\}.
\end{align*}
From \cref{eq:kt} we see that $P_Q(\lambda_1,\lambda_2,\lambda_3,\lambda_4)$ is described by a polynomial number of inequalities, each of which is of the desired form.
To see that the number of integer points in $P_Q(\lambda_1,\lambda_2,\lambda_3,\lambda_4)$ equals $m_Q(\lambda_1,\lambda_2,\lambda_3,\lambda_4)$, note that the fibers of $P_Q(\lambda_1,\lambda_2,\lambda_3,\lambda_4)$ under the projection \[ (\alpha,\beta,\gamma,\delta,p_1,p_2,p_3,p_4) \mapsto (\alpha,\beta,\gamma,\delta) \] are products of Knutson-Tao polytopes.
The number of integral points in $P_Q(\lambda_1,\lambda_2,\lambda_3,\lambda_4)$ can be computed by summing the product of the number of integral points in these Knutson-Tao polytopes over all dominant polynomial weights~$\alpha,\beta,\gamma,\delta$.
By \cref{eq:C vs KT}, the number of integral points in each Knutson-Tao polytope computes the corresponding Clebsch-Gordan coefficient.
Thus we obtain~\eqref{eq:m1234}.

\subsection{The general construction}\label{sec:general}
The above generalizes readily to a general quiver~$Q$ and dimension vector~$\vec n$.

We first define Knutson-Tao polytopes for an arbitrary number of tensor factors (see, e.g.,~\cite{zelevinsky1999littlewood}).
Given dominant weights $\lambda_1,\ldots,\lambda_s$ for~$\GL(n)$, denote by $C(\lambda_1,\ldots,\lambda_s)$ the multiplicity of the trivial representation in the tensor product $V^{\lambda_1}\otimes \cdots\otimes V^{\lambda_s}$.
Then one can define polytopes, which we will denote by~$\KT(\lambda_1,\ldots,\lambda_s)$, such that the number of integral points is equal to~$C(\lambda_1,\lambda_2,\ldots,\lambda_s)$.
The following construction will suffice for our purposes:
The case~$s\leq2$ are trivial, the case $s=3$ is the one originally covered by Knutson-Tao and defined in \cref{eq:kt}, and for~$s>3$ one has
\begin{align*}
  C(\lambda_1,\dots,\lambda_s)
= \sum_\mu C(\mu^*, \lambda_1,\dots,\lambda_{s-2}) C(\mu,\lambda_{s-1},\lambda_s),
\end{align*}
where~$\mu$ runs over all dominant weights, so we can inductively define
\begin{align*}
    \KT(\lambda_1,\ldots,\lambda_s)
= \Bigl\{
  &(\mu,p,q) \in \R^n \op W_n \op W_n \;:\; \mu_1 \geq \dots \geq \mu_n, \\
  &\qquad\qquad p \in \KT(\mu^*, \lambda_1,\dots,\lambda_{s-2}), \; q \in  \KT(\mu,\lambda_{s-1},\lambda_s)
\Bigr\}.
\end{align*}
Inductively, we see that the number of inequalities is polynomial in~$n$ and~$s$, that their right-hand sides depend linearly on $\lambda_1,\dots,\lambda_s$, and that all coefficients are in $\{0,1,-1\}$.

Now consider an arbitrary quiver~$Q$ without cycles and an arbitrary dimesnion vector~$\vec n$.
Using the Cauchy formula in its general form, \cref{eq:cauchy general}, we obtain the following decomposition with respect to~$\GL_Q(\CV)$:
\begin{align*}
  \Sym^*(\mathcal H_Q(\CV))
&= \bigoplus_{\{\mu_a \in D(n_x,n_y) \cap \Z^{n_x}\}_{a : x \to y \in Q_1}} \bigotimes_{a : x \to y \in Q_1} V^{\mu_a}_x \ot V^{d(\mu_a)^*}_y
\\
&= \bigoplus_{\{\mu_a \in D(n_x,n_y) \cap \Z^{n_x}\}_{a : x \to y \in Q_1}}
  \bigotimes_{x \in Q_0}
  \left(
  \bigotimes_{a \in \outarrows(x)} V^{\mu_a}_x
  \otimes
  \bigotimes_{a \in \inarrows(x)} V^{d(\mu_a)^*}_x
  \right),
\end{align*}
where we write $\outarrows(x)$ and $\inarrows(x)$ for the set of arrows starting from~$x$ and arriving at~$x$, respectively, and where for each vertex~$x$ the corresponding factor $\GL(n_x)$ of $\GL_Q(\CV)$ again acts diagonally on all tensor factors with subscript~$x$.
Hence, similarly to~\cref{eq:m1234}, the multiplicity~$m_Q(\vec\lambda)$ of a highest weight~$\vec\lambda=(\lambda_x)_{x\in Q_0}$ can be computed in the following way:
\begin{equation*}
  m_Q(\vec\lambda)
= \sum_{\{\mu_a \in D(n_x,n_y) \cap \Z^{n_x}\}_{a : x \to y \in Q_1}}
  \prod_{x \in Q_0} C(\lambda_x^*, \{ \mu_a \}_{a \in \outarrows(x)}, \{ d(\mu_a)^* \}_{a \in \inarrows(x)}),
\end{equation*}
We thus define the polytope:
\begin{align*}
  P_Q(\vec\lambda)
= \Bigl\{
  &\bigl( (\mu_a\}_{a \in Q_1}, (p_x)_{x \in Q_0} \bigr) \;:\;
  \mu_a \in D(n_x,n_y) \quad \forall a:x\to y \in Q_1, \\
  &p_x \in \KT\bigl(\lambda_x^*, \{ \mu_a \}_{a \in \outarrows(x)}, \{ d(\mu_a)^* \}_{a \in \inarrows(x)}\bigr) \quad \forall x \in Q_0
\Bigr\}
\end{align*}
Note that $P_Q(\vec\lambda)$ can be described by a combinatorial linear program with a number of inequalities that is polynomial in the number of vertices and the number of arrows of the quiver, with right-hand sides depending linearly on $\lambda_1,\dots,\lambda_s$, and with all coefficients in $\{0,1,-1\}$.
It is clear that $m_Q(\vec\lambda)$ equals the number of integral points in this polytope.
Moreover, the combinatorial linear program can itself be generated in strongly polynomial time given~$Q$ and~$\vec\lambda$.
Thus we have proved \cref{thm}.
As explained in the introduction, \cref{cor} as a consequence.

For the class of bipartite quivers considered in \cite{chindris2022membership}, the authors similarly construct polytopes whose number of integer points compute the corresponding multiplicities.
Their construction has a nice geometric description by gluing together just \emph{two} Knutson-Tao polytopes, using Zelevinsky's polyhedral description for higher tensor product multiplicities~\cite{zelevinsky1999littlewood}.

\section{Semi-invariants and complexity of the generic semi-stability problem}\label{sec:semi}
Given a quiver~$Q$ and a dimension vector~$\vec n\in \N^{Q_0}$, let~$\Sigma_Q(\vec n)$ denote the cone generated by the weights of semi-invariant polynomials for this data.
By saturation, $\vec\sigma \in \Sigma_Q(\vec n)$ if there exists a non-zero semi-invariant polynomial on the space~$\CH_Q(\CV)$ of weight~$\vec\sigma$.
Chindris, Collins, and Kline asked if one can give a strongly polynomial time algorithm for the problem of deciding membership in~$\Sigma_Q(\vec n)$, which they called the \emph{generic semi-stability problem}~\cite{chindris2022membership}*{Problem~2}.

Mathematically, the generic semi-stability problem can be seen as a special case of moment cone membership.
Indeed, let us associate to any~$\vec\sigma \in \Z^{\Q_0}$ the dominant weight~$\vec\lambda(\vec\sigma)$ defined by~$\lambda_x = (\sigma_x,\dots,\sigma_x) \in \Z^{n_x}$ for all~$x\in Q_0$.
Then it holds that
\begin{align*}
  \vec\sigma \in \Sigma_Q(\vec n)
\quad\Leftrightarrow\quad
  m_Q(\vec\lambda(\vec\sigma)) > 0
\quad\Leftrightarrow\quad
  P_Q(\vec\lambda(\vec\sigma)) \neq \emptyset.
\end{align*}

However, the existence of the strongly polynomial time algorithm for moment cone membership in \cref{cor} does \emph{not} imply the existence of a (strongly or not) polynomial time algorithm for the generic semi-stability problem.
Indeed, the input data for the generic semi-stability problem naturally consists of a description of the quiver~$Q$ along with two list of integers of length~$\lvert Q_0 \rvert$ describing the dimension vector~$\vec n \in \N^{Q_0}$ and weight~$\vec \sigma \in \Z^{Q_0}$, respectively.
Thus the \emph{number of} input numbers for this problem is independent of the dimensions~$n_x$ (unlike for the moment cone membership problem).
However, the polytope~$P_Q(\vec\lambda)$ constructed above has a description complexity that grows at least linearly with the dimensions~$n_x$.
Thus it cannot be constructed in strongly polynomial time from the input data.
In fact, when the the dimensions~$n_x$ are specified in binary, as is usual in computational complexity theory, one cannot even generate~$P_Q(\vec\lambda)$ in ordinary (non-strongly) polynomial time from the input data.
We conclude that our result for moment cone membership does \emph{not} imply a (strongly or not) polynomial time algorithm for the generic semi-stability problem.
Since our construction is very similar to the one of~\cite{chindris2022membership}, we believe that the existence of strongly polynomial time algorithms for the generic semi-stability problem as stated in~\cite{chindris2022membership} is still open.%
\footnote{This also applies to the special case of star quivers, recently discussed in \cite{franks2023ipca}.}

In the remainder we show that at least for certain special quivers, the generic semi-stability problem can be decided in polynomial time.
Our first example is the quiver~$Q$ in~\eqref{eq:quiver}:
\begin{equation*}
  1 \rightarrow 3 \leftarrow 2
\end{equation*}
Given a dimension vector $\vec n=[n_1,n_2,n_3] \in\N^3$ and a weight $\vec\sigma=[\sigma_1,\sigma_2,\sigma_3]\in \Z^3$, we would like to understand when $\vec\sigma \in \Sigma_Q(\bm n)$.
If $n_1=n_2=n_3$, then this is the case if and only if
\begin{align*}
  \sigma_1 \geq 0, \quad \sigma_2 \geq 0, \quad \sigma_1 + \sigma_2 + \sigma_3 = 0.
\end{align*}
If $n_1=n_3\neq n_2$ then $\vec\sigma \in \Sigma_Q(\bm n)$ if and only if~$\sigma_1 + \sigma_3 = 0$ and~$\sigma_2 = 0$.
Similarly, if~$n_1\neq n_3=n_2$ then~$\vec\sigma \in \Sigma_Q(\bm n)$ if and only if~$\sigma_1 = 0$ and~$\sigma_2 + \sigma_3 = 0$.
Finally, if~$n_1 + n_2 = n_3$ then~$\vec\sigma \in \Sigma_Q(\bm n)$ if and only if~$\sigma_1 = \sigma_2 = -\sigma_3$.
In all other cases, $\Sigma_Q(\bm n) = \{0\}$.
We see that for this quiver, the generic semi-stability problem can be decided in strongly polynomial time.

Next, we discuss the quiver in~\eqref{eq:quiver4}:
\begin{align*}
  &\qquad ~2 \qquad \\
  &\nearrow ~\qquad \searrow \\
  1 & \qquad\qquad\quad 4 \; , \\
  &\searrow ~\qquad \nearrow \\
  &\qquad ~3 \qquad
\end{align*}
Let~$\bm n=[n_1,n_2,n_3,n_4]\in \N^4$ and let~${\bm \sigma}=[\sigma_1,\sigma_2,\sigma_3,\sigma_4] \in \Z^4$.
Again we would like to understand when $\vec\sigma \in \Sigma_Q(\bm n)$.
A necessary condition is that $n_1 \sigma_1 + \dots + n_4 \sigma_4 = 0$, thus the cone is at most three-dimensional.
It can be computed by using the Cauchy formula in \cref{eq:cauchy general}.
The upshot is as follows:
\begin{align*}
  \Sigma_Q(\vec n) = \mathfrak C(\vec n) \cap  \{ n_1 \sigma_1 + \dots + n_4 \sigma_4 = 0 \},
\end{align*}
where $\mathfrak C(\vec n)$ ranges over a \emph{finite} number of polyhedral convex cones in~$\R^4$ as~$\vec n$ is varied over all (infinitely many) dimension vectors.
Furthermore, given~$\vec n$, one can by using a bounded number of elementary arithmetic operations compute defining inequalities of the corresponding cone~$\mathfrak C(\vec n)$.
Thus the generic semi-stability problem can again be decided in strongly polynomial time.
Since there is a large (but finite) number of cases to consider we only discuss a few examples.
When~$\vec n = [n,n,n,n]$ then~$\Sigma_Q(\bm n)$ is the three-dimensional cone generated by the vectors~$(1,-1,0,0)$, $(1,0,-1,0)$, $(0,1,0,-1)$, $(0,0,1,-1)$, corresponding to the determinants for the four arrows.
Thus,~$\vec\sigma \in \Sigma_Q(\bm n)$ if and only if
\begin{align*}
  \sigma_1 \geq 0, \quad
  \sigma_1 + \sigma_2 \geq 0, \quad
  \sigma_1 + \sigma_3 \geq 0, \quad
  \sigma_1 + \sigma_2 + \sigma_3 \geq 0, \quad
  \sigma_1 + \sigma_2 + \sigma_3 + \sigma_4 = 0.
\end{align*}
When the dimensions are increasing by at most one along each of the two paths~$1\to2\to4$ and~$1\to3\to4$, then~$\Sigma_Q(\bm n)$ is again three-dimensional.
For $\vec n = [n,n,n,n+1]$ it is given by
\begin{align*}
  \sigma_1 + \sigma_2 + \sigma_4 \geq 0, \quad
  \sigma_1 + \sigma_3 + \sigma_4 \geq 0, \quad
  \sigma_4 \leq 0, \quad
  n \sigma_1 + n \sigma_2 + n \sigma_3 + (n+1) \sigma_4 = 0;
\end{align*}
for $\vec n = [n,n+1,n,n+1]$ it is described by
\begin{align*}
  \sigma_1 + \sigma_3 \geq 0, \quad 
  \sigma_1 + \sigma_2 + \sigma_4 \geq 0, \quad
  \sigma_2 \geq 0, \quad
  n \sigma_1 + (n+1) \sigma_2 + n \sigma_3 + (n+1) \sigma_4 = 0;
\end{align*}
and similarly for $\vec n = [n,n,n+1,n+1]$, using symmetry;
finally, for $\vec n = [n,n+1,n+1,n+1]$, the cone is given by
\begin{align*}
  \sigma_1 \geq 0, \quad
  \sigma_2 \geq 0, \quad
  \sigma_3 \geq 0, \quad
  n \sigma_1 + (n+1) \sigma_2 + (n+1) \sigma_3 + (n+1) \sigma_4 = 0.
\end{align*}
When the dimensions are increasing by more than one along the two paths (i.e., $n_4 \geq n_1 + 2$) then there are again several cases to consider.
For example, the case that~$n_1 < n_2 < n_4$, $n_1 < n_3 < n_4$, and~$n_4 \leq 2 n_1$ is particularly interesting, since now $\mathfrak C(\vec n)$ depends also on the cardinality of a certain set of remainders of Euclidean divisions of integers in~$\{n_j, n_j - n_k\}$.

\section*{Acknowledgments}
We would like to thank Calin Chindris, Brett Collins, and Daniel Kline for valuable discussions about their preprint~\cite{chindris2022membership}.
We also acknowledge discussions with Cole Franks and Visu Makam on an earlier version of this note.

MW acknowledges support by the European Union (ERC, SYMOPTIC, 101040907), by the Deutsche For\-schungs\-ge\-mein\-schaft (DFG, German Research Foundation) under Germany's Excellence Strategy - EXC\ 2092\ CASA - 390781972, by the German Federal Ministry of Education and Research (BMBF, project QuBRA), and by the Dutch Research Council (NWO, grant OCENW.KLEIN.267).
Views and opinions expressed are those of the author(s) only and do not necessarily reflect those of the European Union or the European Research Council Executive Agency.
Neither the European Union nor the granting authority can be held responsible for them.

\bibliographystyle{amsalpha}
\bibliography{quiver-polytopes}

\newcommand{\etalchar}[1]{$^{#1}$}
\providecommand{\bysame}{\leavevmode\hbox to3em{\hrulefill}\thinspace}
\providecommand{\MR}{\relax\ifhmode\unskip\space\fi MR }
\providecommand{\MRhref}[2]{%
  \href{http://www.ams.org/mathscinet-getitem?mr=#1}{#2}
}
\providecommand{\href}[2]{#2}
\begin{thebibliography}{BCMW17}

\bibitem[BCMW17]{burgisser2017membership}
Peter Burgisser, Matthias Christandl, Ketan~D Mulmuley, and Michael Walter,
  \emph{Membership in moment polytopes is in {NP} and {coNP}}, SIAM Journal on
  Computing \textbf{46} (2017), no.~3, 972--991.

\bibitem[BFG{\etalchar{+}}19]{burgisser2019towards}
Peter B{\"u}rgisser, Cole Franks, Ankit Garg, Rafael Oliveira, Michael Walter,
  and Avi Wigderson, \emph{Towards a theory of non-commutative optimization:
  Geodesic 1st and 2nd order methods for moment maps and polytopes}, 2019 IEEE
  60th Annual Symposium on Foundations of Computer Science (FOCS), IEEE, 2019,
  pp.~845--861.

\bibitem[BS00]{berenstein2000coadjoint}
Arkady Berenstein and Reyer Sjamaar, \emph{Coadjoint orbits, moment polytopes,
  and the {H}ilbert--{M}umford criterion}, Journal of the American Mathematical
  Society \textbf{13} (2000), no.~2, 433--466.

\bibitem[Buc00]{buch2000saturation}
Anders~Skovsted Buch, \emph{The saturation conjecture (after {A}.~{K}nutson and
  {T}.~{T}ao)}, Enseign. Math. (2) \textbf{46} (2000), no.~1/2, 43--60, with an
  appendix by William Fulton.

\bibitem[BVW19]{baldoni2019horn}
Velleda Baldoni, Mich{\`e}le Vergne, and Michael Walter, \emph{Horn conditions
  for quiver subrepresentations and the moment map}, arXiv preprint
  arXiv:1901.07194 (2019), to appear in Pure Appl. Math. Q.

\bibitem[CCK22]{chindris2022membership}
Calin Chindris, Brett Collins, and Daniel Kline, \emph{Membership in moment
  cones, quiver semi-invariants, and generic semi-stability for bipartite
  quivers}, arXiv preprint arXiv:2211.01990 (2022), Nov 2022.

\bibitem[DW00]{derksen2000semi}
Harm Derksen and Jerzy Weyman, \emph{Semi-invariants of quivers and saturation
  for {L}ittlewood--{R}ichardson coefficients}, Journal of the American
  Mathematical Society \textbf{13} (2000), no.~3, 467--479.

\bibitem[FM23]{franks2023ipca}
Cole Franks and Visu Makam, \emph{{iPCA} and stability of star quivers}, arXiv
  preprint arXiv:2302.09658 (2023).

\bibitem[Ful97]{fulton1997young}
William Fulton, \emph{Young tableaux}, Cambridge University Press, 1997.

\bibitem[GLS12]{grotschel2012geometric}
Martin Gr{\"o}tschel, L{\'a}szl{\'o} Lov{\'a}sz, and Alexander Schrijver,
  \emph{Geometric algorithms and combinatorial optimization}, vol.~2, Springer
  Science \& Business Media, 2012.

\bibitem[KT99]{knutson1999honeycomb}
Allen Knutson and Terence Tao, \emph{The honeycomb model of
  $\operatorname{GL}_n(\mathbbm{C})$ tensor products {I}: {P}roof of the
  saturation conjecture}, Journal of the American Mathematical Society
  \textbf{12} (1999), no.~4, 1055--1090.

\bibitem[MNS12]{mulmuley2012geometric}
Ketan~D. Mulmuley, Hariharan Narayanan, and Milind Sohoni, \emph{Geometric
  complexity theory {III}: on deciding nonvanishing of a
  {L}ittlewood--{R}ichardson coefficient}, J. Algebraic Combin. \textbf{36}
  (2012), 103--110, arXiv:cs/0501076.

\bibitem[Res10]{ressayre2010geometric}
Nicolas Ressayre, \emph{Geometric invariant theory and the generalized
  eigenvalue problem}, Inventiones mathematicae \textbf{180} (2010), no.~2,
  389--441.

\bibitem[Tar86]{tardos1986strongly}
{\'E}va Tardos, \emph{A strongly polynomial algorithm to solve combinatorial
  linear programs}, Operations Research \textbf{34} (1986), no.~2, 250--256.

\bibitem[VW17]{vergne2017inequalities}
Mich{\`e}le Vergne and Michael Walter, \emph{Inequalities for moment cones of
  finite-dimensional representations}, Journal of Symplectic Geometry
  \textbf{15} (2017), no.~4, 1209--1250.

\bibitem[Zel99]{zelevinsky1999littlewood}
Andrei Zelevinsky, \emph{{L}ittlewood--{R}ichardson semigroups}, New
  Perspectives in Algebraic Combinatorics, Cambridge University Press, 1999,
  pp.~337--345.

\end{thebibliography}

\end{document}